\documentclass[reprint,superscriptaddress,amsmath,amssymb,aip, apl,nolongbibliography]{revtex4-2}

\usepackage{pifont}
\usepackage{graphicx}% Include figure files
\usepackage{bm}% bold math
\usepackage{xcolor}
\usepackage[colorlinks=true,citecolor=blue]{hyperref}
\usepackage{natbib}
\usepackage{hyperref}

\newcommand{\Rmnum}[1]{\expandafter\@slowromancap\romannumeral #1@}
\makeatother

\begin{abstract}
Exceptional points are complex branching singularities of non-Hermitian bands that have lately attracted considerable interest, particularly in non-Hermitian photonics. In this article, we review some recent developments in non-Hermitian photonic platforms such as waveguides, photonic crystals, Fabry-P\'{e}rot resonators and plasmonic systems, and suggest how optical non-linearities and exceptional bound states can significantly impact the development of non-Hermitian photonics in the near future.
\end{abstract}

\begin{document}

\title{Exceptional points in non-Hermitian Photonics: Applications and Recent Developments}

\author{Haiyu Meng}
\email{h.y.meng@hnu.edu.cn}
%\thanks{Authors to whom correspondence should be addressed: H.Y.Meng@hnu.edu.cn, yeesin\_ang@sutd.edu.sg and phylch@nus.edu.sg}
\affiliation{School of Physics and Optoelectronics, Xiangtan University, Xiangtan 411100, China}
\affiliation{Department of Physics, National University of Singapore, Singapore 117542}

\author{Yee Sin Ang}
\email{yeesin\_ang@sutd.edu.sg}
%\thanks{Authors to whom correspondence should be addressed: H.Y.Meng@hnu.edu.cn, yeesin\_ang@sutd.edu.sg and phylch@nus.edu.sg}
\affiliation{Science, Mathematics and Technology (SMT), Singapore University of Technology and Design, Singapore 497372}

\author{Ching Hua Lee}
\email{phylch@nus.edu.sg}
%\thanks{Authors to whom correspondence should be addressed: H.Y.Meng@hnu.edu.cn, yeesin\_ang@sutd.edu.sg and phylch@nus.edu.sg}
\affiliation{Department of Physics, National University of Singapore, Singapore 117542}
\affiliation{Joint School of National University of Singapore and Tianjin University, International Campus of Tianjin University, Binhai New City, Fuzhou 350207, China}

\maketitle

%\begin{Introduction}

%\end{Introduction}

\section{Introduction}

For much of the twentieth century, most solid state research have focused on Hermitian quantum mechanical systems, where the Hermiticity of the Hamiltonian was deemed as a necessary condition for stability. This, however, have changed significantly in the recent years, with the ascendancy of metamaterial systems  such as photonic\cite{regensburger2012parity,miri2019exceptional,li2023exceptional,wang2021coherent,ozdemir2019parity,zhen2015spawning,lee2009observation,cao2015dielectric,weidemann2020topological,song2020two,zhu2020photonic,gbur2018introduction}, acoustic\cite{zhang2021observation,zhang2021acoustic,gao2022non}, electrical\cite{helbig2020generalized,hofmann2020reciprocal,liu2021non,zou2021observation,stegmaier2021topological,ezawa2021non,zhang2021observation,shang2022experimental,zhang2022observation,zhang2023electrical} and mechanical\cite{Brandenbourger2019,ghatak2020observation} arrays that naturally incorporate gain and loss, as well as non-reciprocity. Non-Hermitian photonic systems, in particular, have arguably seen the most vibrant developments of late, with optical loss intrinsically present in most optical media and optical gain routinely occurring in lasers. Indeed, the enormous design flexibility, ease-of-use of electromagnetic simulation software and maturing fabrication technology have made photonic systems extremely versatile platforms for manipulating and controlling light in ways not straightforwardly attainable in electronic crystals. 

%Such characteristic can be realized in photonic systems using optical gain or loss media, thus enabling the realization of unconventional optical phenomena arising from the interplay between gain and loss.

\begin{figure}[h]
\includegraphics[width = 1\linewidth]{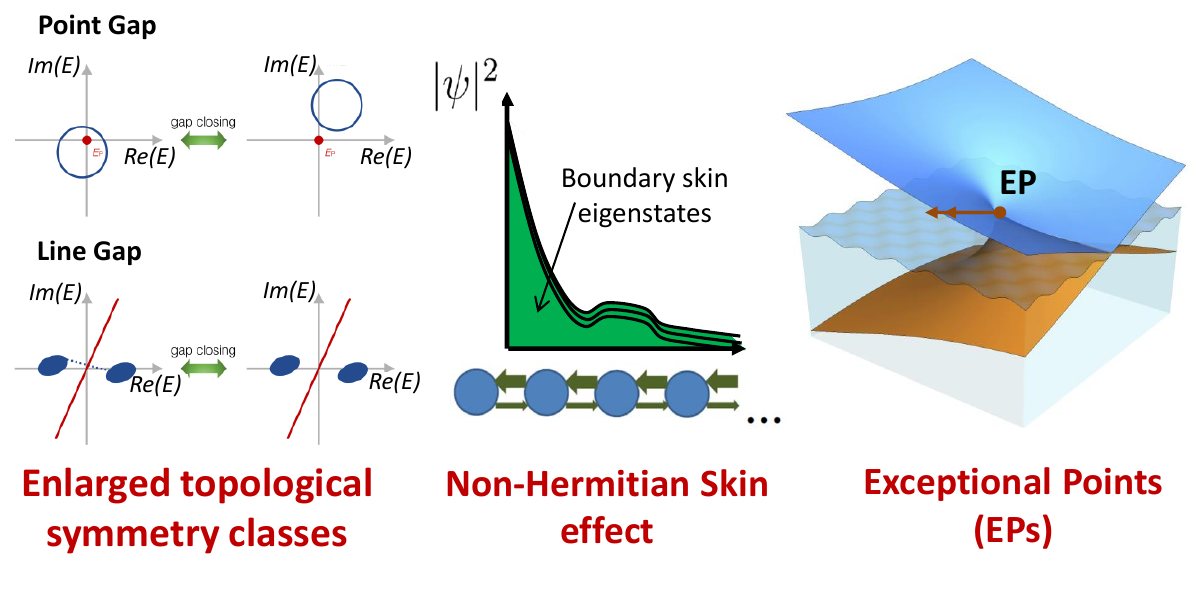}
\caption{Three consequences of complex non-Hermitian spectra. (Left) On the complex spectral plane, eigenbands (blue) may be separated by line gaps that are direct generalizations of usual Hermitian band gaps, or may non-trivially encircle nonzero areas and form so-called point gaps, thereby leading to much richer symmetry-protected topological states  [Figure adapted from Ref.~\onlinecite{kawabata2019symmetry}]. (Center) Spectral loops around point gaps generically collapse when physical boundaries are introduced, leading to the boundary accumulation of all states with an exponentially-decaying envelope (green) i.e. the well-known non-Hermitian skin effect. (Right) Eigenbands in the complex plane can also exhibit branch cuts terminating at exceptional points (EPs), where the dispersion diverges like a polynomial-order singularity and the eigenstates (brown) coalesce. The latter also leads to enigmatic exceptional bound states\cite{lee2022exceptional,zou2023experimental} in the incomplete subspace spanned by one of the bands.  }
\label{fig:intro}
\end{figure}

Qualitatively new phenomena with no Hermitian analogs originate from the complex nature of non-Hermitian spectra, which lie on the 2D complex plane instead of the 1D real line. Here, the spectra may refer to not just the eigenenergies of a Hamiltonian, but can also be the eigenvalues of a network Laplacian\cite{helbig2020generalized}, scattering matrix\cite{franca2022non} or even reduced density matrices\cite{lee2022exceptional}.  As illustrated in Fig.\ref{fig:intro}(Left), non-Hermitian bands can either exist as ``blobs'' in the complex plane that are separable by line gaps, which directly generalize usual Hermitian gaps, or they can also enclose nonzero areas forming so-called point gaps. Together, the variety of topologically inequivalent gaps augment the existing classification of symmetry-protected topological states\cite{kawabata2019symmetry,okuma2020topological}. When lattice translation symmetry is broken by boundaries or imperfections, point gapped spectra generically collapse into their interior non-perturbatively\cite{xiong2018does, lee2019anatomy, yokomizo2019non}, leading to the classic non-Hermitian skin effect characterized by extensive boundary state localization\cite{yao2018edge,lin2023topological,zhang2022review,song2019non,edvardsson2019non,borgnia2020non,Wanjura2020,wanjura2021correspondence,xue2021simple,longhi2022self,xue2022non,lee2020ultrafast,budich2020sensor,guo2021exact,li2021impurity,li2020critical,yokomizo2021scaling,liu2020helical,kawabata2020higher,okugawa2020second,fu2021nonH,sun2021geometric,lee2021many,bhargava2021non,li2021quantized,schindler2021dislocation,panigrahi2022non,manna2022inner,zhang2022universal,yang2022designing,oztas2019schrieffer,zhu2021delocalization,zhang2021tidal,lee2020unraveling,longhi2022non,shen2022non,cheng2022competition,li2022non,okuma2022non,jiang2022filling,gu2022transient,tai2023zoology,lei2023mathcal,qin2023kinked,okuma2023non,yang2023percolation,cao2023many,jiang2023dimensional,kawabata2023entanglement,qin2023universal} [Fig.\ref{fig:intro}(Center)].

\begin{figure*}%[ht]
\centering
\includegraphics[width = 0.92\linewidth]{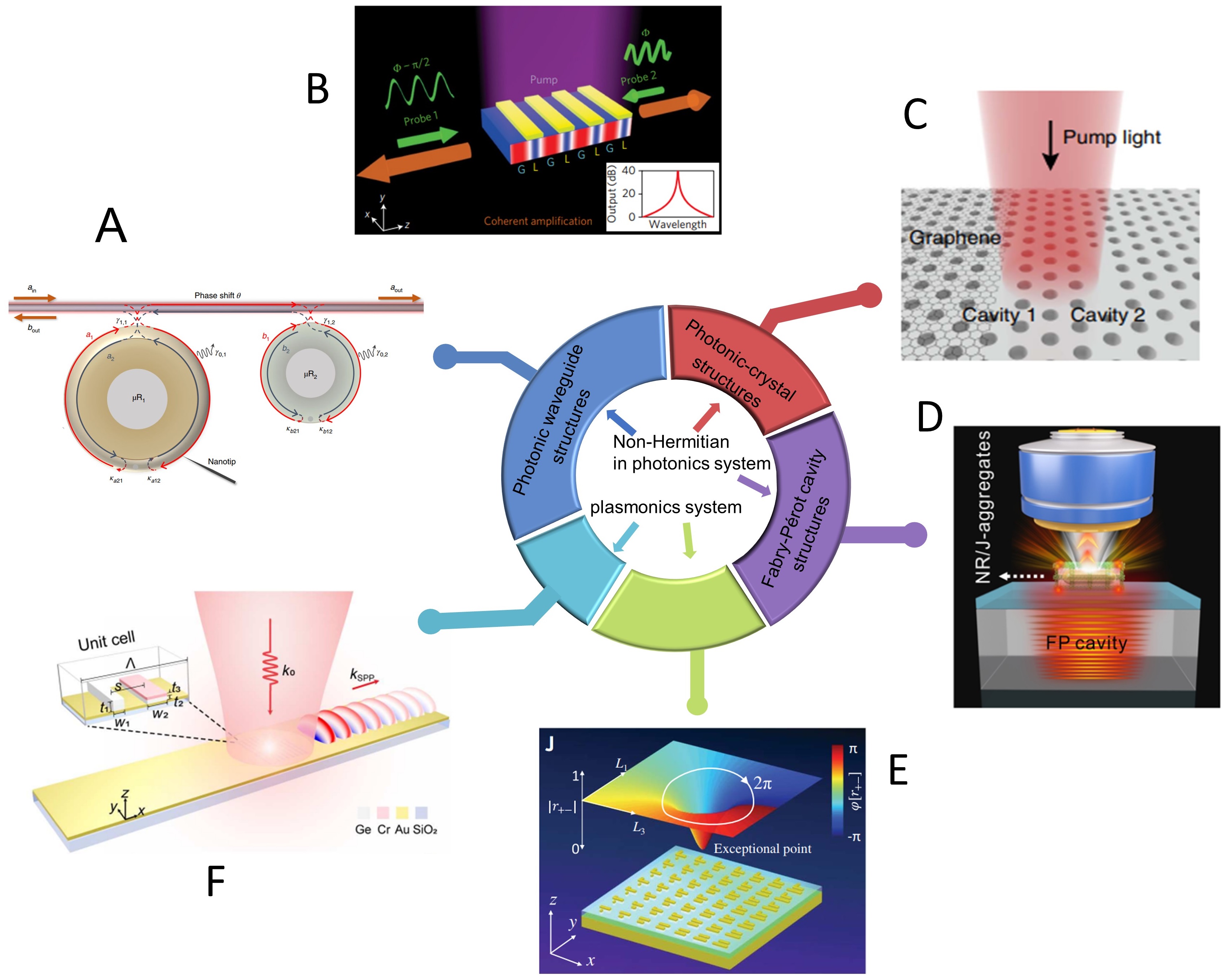}
\caption{Some key photonics and plasmonics experiments based on non-Hermitian EPs, to be mostly elaborated further in the later sections. (A) Electromagnetically-induced transparency achieved by tuning the coupling between whispering-gallery-mode micro-resonators to an EP\cite{wang2020electromagnetically}. (B) Coherent perfect absorber-laser achieved by realizing both lasing and anti-lasing in a single cavity, made possible with balanced gain/loss at an EP\cite{wong2016lasing}. (C) An EP phase transition in a photonic crystal is observed by tuning the area in which it is covered by a graphene sheet\cite{kim2016direct}. (D) The EP transition to the strong coupling regime of a single-exciton with localized plasmon modes is achieved using a leaky Fabry–Perot cavity\cite{li2023highly}. (E) In a planar chiral plasmonic metasurface, the topological encircling of an EP in parameter space is used to realize a $2\pi$ phase in a specific reflected polarization channel, representing an additional avenue in optical phase engineering\cite{song2021plasmonic}. (F) A spatially modulated dielectric metagrating with effective real-imaginary permittivity phase shift results inu nidirectional excitation and reflection of surface plasmon polaritons when tuned near an EP\cite{xu2023subwavelength}.}
\label{fig:1}
\end{figure*}

Even more interestingly, the possibility of non-diagonalizability of generic non-Hermitian operators leads to geometric defectiveness at the eigenspace whenever its Jordan form becomes non-trivial. These instances, known as the famed exceptional points (EPs), are also the branch points of the band structure [Fig.\ref{fig:intro}(Right)]. The technological potential of EPs broadly stems for their following characteristics:
\begin{enumerate}
\item As a branch point, an EP can possess multi-valued bands around it, a property that enables mode switching\cite{doppler2016dynamically} or topological energy transfer\cite{xu2016topological} applications.
\item The spectra bifurcation around a second-order EP can give rise to a $\mathcal{PT}$-symmetry transition from real to complex eigenvalues\cite{ashida2020non,schomerus2013scattering,ni2018pt}, leading to a very precise means to control the asymptotic system dynamics. 
\item As an algebraic singularity, an $N$-th order EP possesses a divergent dispersion of the form $E(k)\sim k^{1/N}$ near it, which hence leads to extreme sensitivity as the parameter $k$ is varied.  
\item The geometric defectiveness at an EP can give rise to specially growing eigenstates\cite{longhi2018exceptional}, with the  coalescence of EP eigenstates allowing the system to take multiple roles simultaneously i.e. laser and coherent perfect absorber\cite{chong2011p}.
\item More esoterically, the defectiveness of a momentum-space EP protects special exceptional bound states\cite{lee2022exceptional, zou2023experimental} when confined to a particular band and spatial region, leading to a novel type of robust band with no Hermitian analog, as further prognosticated in the final section.
\end{enumerate}

\section{Exceptional points in photonic systems}

Unlike in quantum materials, EPs have already been demonstrated and even harnessed in a wide array of photonics and plasmonics platforms. A sampling of some key experimental platforms is shown in Fig.~\ref{fig:1} and elaborated in the subsequent sections. They include whispering-gallery-mode micro-resonators (Fig.~\ref{fig:1}A)\cite {wang2020electromagnetically}, coherent perfect absorber-laser waveguides (Fig.~\ref{fig:1}B)\cite {wong2016lasing}, photonic crystal cavities with a graphene control layer (Fig.~\ref{fig:1}C) \cite{kim2016direct}, Fabry–Perot cavities (Fig.~\ref{fig:1}D)\cite {li2023highly}, plasmonic metasurfaces (Fig.~\ref{fig:1}E)\cite{song2021plasmonic}, and plasmonic metagratings (Fig~\ref{fig:1}F) \cite{xu2023subwavelength}.
Beyond acting as definitive experimental platforms for demonstrating fundamental physics, photonic EP realizations  open a new avenue towards the design of functional optical components and devices such as highly sensitive sensors, lasers and optical modulators \cite{chen2017exceptional,wiersig2014enhancing,ozdemir2019parity,miri2019exceptional,hodaei2017enhanced,liertzer2012pump,heiss2012physics,lin2017line,budich2020non,yuan2023non,feng2017non,feng2013experimental,li2023exceptional,ma2022nonlinear}.

%How can photonic systems effectively facilitate the exploration of non-Hermitian physics? Mathematically, their resemblance to quantum systems and the flexibility in parameter tuning and modulation make photonic systems an exceptional platform for a wide range of photonic device applications. 

Why are photonic systems so suitable for exploring non-Hermitian physics? Other than their great tunability, photonic media are amenable to the introduction of artificially constructed gain and loss. The equivalence between the optical paraxial wave equation and the Schr\"{o}dinger equation in quantum mechanics \cite{el2007theory,longhi2018parity,feng2017non,gupta2020parity} provides a versatile optical sandbox for probing $\mathcal{PT}$-symmetry physics \cite{ozdemir2019parity,ruter2010observation}. Consider an optical medium with complex refractive index $n(\vec r)=n_r(\vec r)+in_i(\vec r)$. $\mathcal{PT}$-symmetry requires that the refractive index satisfies $n(\vec r)= n^*(-\vec{r})$, which corresponds to $n_r(\vec{r})=n_r(-\vec{r})$ and $n_i(\vec{r})=-n_i(-\vec{r})$. Fortunately, this is implementable in active and dissipative optical media \cite{yan2023advances}, thereby enabling interesting non-Hermitian photonics device applications \cite{li2023exceptional,miri2016nonlinearity,ozawa2019topological,ozdemir2019parity}.

%% CH: This paragraph is now incorporated into the other paragraphs
%In $\mathcal{PT}$-symmetric photonic structures, despite their non-Hermitian nature, the structures exhibit intriguing properties in their spectral behavior. The eigenvalue spectra appear entirely real, defying the typical association of non-Hermitian systems with complex domains or complex conjugate eigenvalue pairs \cite{ashida2020non,schomerus2013scattering,ni2018pt}. \textcolor{green}{In the scenario of entire real eigenvalue, remarkably, the eigenvalues also maintain $\mathcal{PT}$ symmetry, while, complex conjugate eigenvalues emerge, indicating the breakdown of $\mathcal{PT}$ symmetry.} \textcolor{red}{[YS: What 'fomer' or 'latter' are you referring to? Maybe this sentence is not needed.]}\textcolor{green}{the former and latter cases refer to eigenvalue spectra appearing entirely real, and complex conjugate pairs, respectively.} This behavior results in the optical intensities predominantly concentrated either within the gain or loss regions of the system. To further understand these intriguing phenomena, researchers have explored various optical structures designed to realize $\mathcal{PT}$ symmetry. In the following sections, we review some representative structures commonly used in exploring the non-Hermitian physics in the photonics platform. 

\begin{figure}[h]
\includegraphics[width = 1\linewidth]{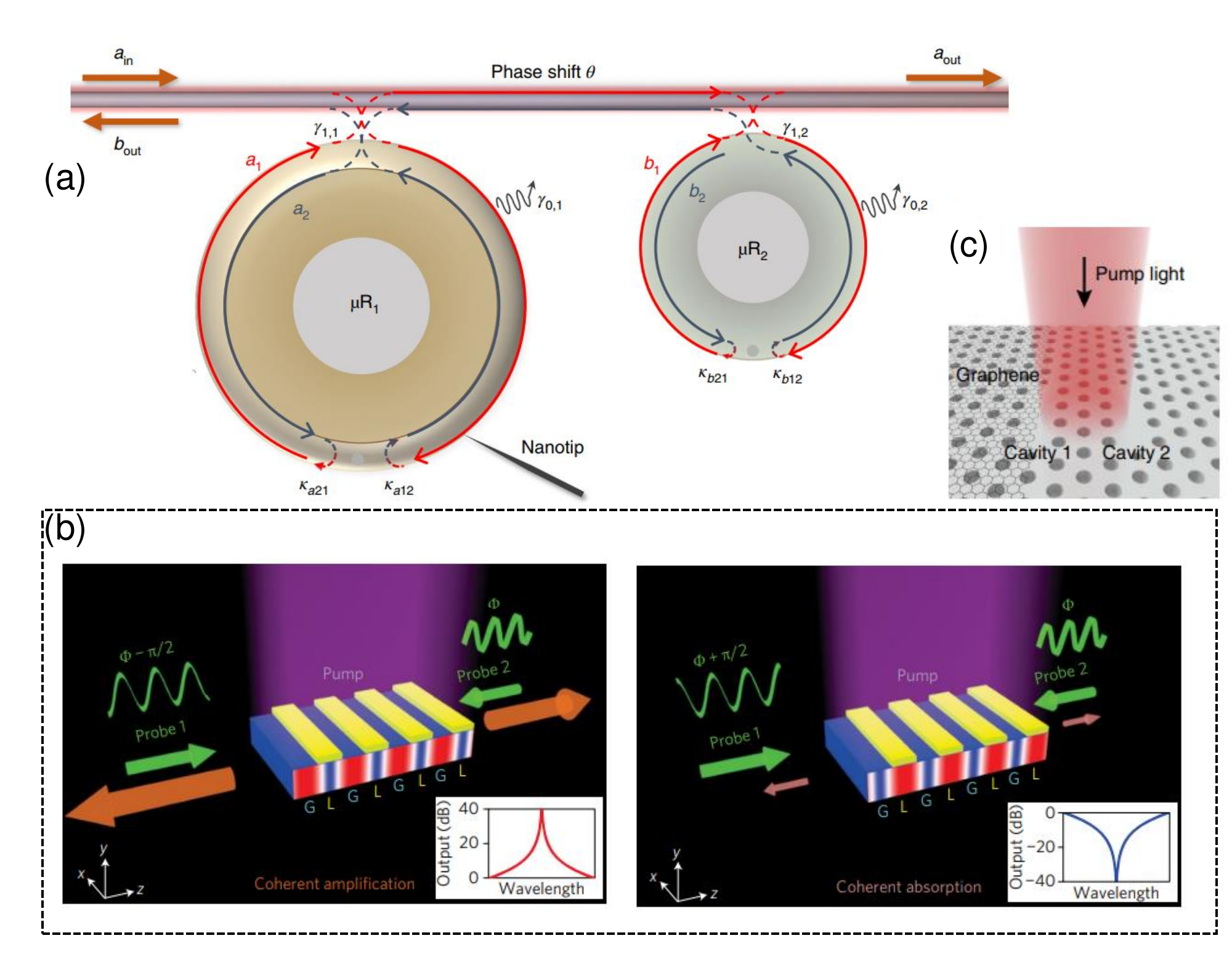}
\caption{(a) In Ref.~\onlinecite{wang2020electromagnetically}, whispering-gallery-mode micro-resonators $\mu R_1$ and $\mu R_2$ are coupled via a fibre taper with a phase shift of $\theta$ such that they exhibit electromagnetically-induced transparency (EIT) when their coupling matrix is tuned to an EP. (b) In Ref.~\onlinecite{wong2016lasing}, a $\mathcal{PT}$-symmetric coherent perfect absorber-laser is realized with a semiconductor InGaAsP/InP gain waveguide coupled to periodic Cr/Ge loss structures (yellow). This coupling creates pure gain (G, blue) and loss (L, red) spatial modulations set at half the wavelength of the guided light. By adjusting the phase offsets $\pm \pi/2$ of the incoming probe beams (green), Bragg interference can confine the electric fields to either the gain (left) or the loss (right) regions, as evident in the output dB plots, thereby making both lasing and anti-lasing eigenmodes possible within the same cavity.
(c) In Ref.~\onlinecite{kim2016direct}, a coupled photonic crystal nanolaser system is effectively tuned to an EP by externally adjusting the extent by which coupled gain cavities 1 and 2 are covered by a lossy graphene sheet.  }
\label{fig:2}
\end{figure}

\subsection{Exceptional points in waveguide structures}
The realm of integrated photonic platforms has yielded noteworthy progress in the realization of EPs and real-complex spectral transitions\cite{liu2022complex,guo2009observation,arkhipov2020quantum,dong2021exceptional,hlushchenko2021multimode,ghosh2016exceptional,wang2023experimental,roccati2022exotic,roccati2023hermitian}. A representative example of such progress is the use of coupled optical waveguides that are carefully engineered to maintain the balanced gain and loss\cite{ruter2010observation}. The integration of gain and loss elements within photonic platforms has enabled efficient control between absorbing and scattering optical modes\cite{wang2020electromagnetically,ozdemir2019parity,li2023exceptional,zhao2015robust}. A particularly prevalent approach  involves the use of whispering-gallery-mode microresonators\cite{yu2021whispering,peng2016chiral,miri2019exceptional,chen2017exceptional,wang2020petermann,lafalce2019robust,hu2021non}, where an EP can be achieved by adjusting the gain–loss contrast and intermodal coupling strength\cite{peng2014loss,peng2014parity}, or by precisely manipulating backscattering and refractive index distribution in one resonator\cite{peng2016chiral,miao2016orbital}. By tuning a whispering-gallery-mode resonator to an EP, the chiral symmetry of clockwise and counterclockwise light propagation can be disrupted, leading to unidirectional invisibility, enhanced sensitivity and mode switching\cite{wang2020electromagnetically}.

One interesting application of EPs in photonics is electromagnetically induced transparency (EIT), as theoretically proposed and experimentally demonstrated in\cite{wang2020electromagnetically} [see Fig.~\ref{fig:2}a]. Originating from destructive interference, EIT is accompanied by drastically reduced optical group velocity associated with applications in optical storage and optical quantum memory. Typically, the realization of EIT in conventional Hermitian structures requires precise tuning and is pervious to noise and thermal fluctuations. Ref.~\onlinecite{wang2020electromagnetically} circumvents these challenges by indirectly coupling whispering-gallery-mode resonators with intrinsic losses, such that the system can be relatively reliably tuned to an EP where there exists only a single merged eigenstate of a fixed chirality of propagation. 
%The introduction of chiral EPs into non-Hermitian systems relaxes such limitation and offers a greater design flexibility of the optical devices. By adjusting the parameters within non-Hermitian systems, such as the ratios of gain and loss, it becomes possible to manipulate the properties and locations of EPs, thus greatly improving the practicality of EIT-based optical devices. 

EPs in waveguide structures can also provide an important building block towards single-mode lasing \cite{zhu2019laser,feng2014single}, a goal that has long been pursued in photonic systems for stable and high-quality laser output. For instance, in $\mathcal{PT}$-symmetric lasers with balanced gain and loss in a $\mathcal{PT}$-symmetric configuration, only a single mode may lase prominently due to the very large discrepancies between the lasing thresholds of different modes near an EP\cite{el2018non,miri2012large}. The infusion of topology further adds the topological stability into the interplay, such as in non-Hermitian gauged topological lasers\cite{longhi2018non}. As active devices, lasers are also intrinsically nonlinear, with the non-linearity inducing interactions between different wave components. This has been explored both theoretically and experimentally in \cite{ezawa2022nonlinear} and\cite{dai2023non}. Moreover, devices based on photonic waveguide structures can simultaneously act as lasers and a coherent perfect absorber at the EPs [see Fig.~\ref{fig:2}b] due to the coalescence of pairs of poles and zeroes in the spectral structure of the scattering matrix. In addition to relatively well-established $\mathcal{PT}$-symmetric non-Hermitian systems, anti-$\mathcal{PT}$-symmetric non-Hermitian setups has also been recently demonstrated in single fiber \cite{bergman2021observation} and subwavelength gratings \cite{liu2022chip}, thus further expanding the horizons of non-Hermitian photonics in waveguide structures.  

\subsection{Exceptional points in photonic crystal structures} 

Photonic crystals are extended optical media with lattice translation symmetry, such that lattice momentum can also enter as an additional parameter degree of freedom. This has made photonic crystals interesting candidates for supporting various non-Hermitian phenomena such as the non-Hermitian skin effect\cite{fang2022geometry,zhong2021nontrivial,wu2022direction,yokomizo2022non,zhu2023photonic}. In the context of EPs, the defectiveness in the momentum-space eigenstates can have drastic implications in the presence of real-space boundaries\cite{lee2022exceptional,zou2023experimental}, as further discussed in the Outlook section. Photonic EPs have been demonstrated in various pioneering studies on $\mathcal{PT}$-symmetric photonic crystals \cite{zhen2015spawning,kaminski2017control,cheng2023realization,zhou2018observation, bykov2018cross,wang2022special}, where the non-Hermiticity can arise due to out-of-plane radiation losses caused by the finite thickness of the dielectric slab \cite{zhen2015spawning,kaminski2017control}. 

%Experimental realization of EPs in coupled photonic crystal nanolasers utilizing the material graphene \cite{kim2016direct} has been recently demonstrated. 
One interesting realization of EPs in coupled photonic crystal nanolasers involves the use of graphene\cite{kim2016direct}, where the EP can be tuned by adjusting the area by which one photonic crystal cavity is covered by graphene. Graphene plays a crucial role in controlling gain contrast, especially in closely spaced wavelength-scale coupled cavities [Fig.~\ref{fig:2}c]. Conventional optical pumping methods often struggle to provide asymmetric gain \cite{schuller2010plasmonics}. However, the use of graphene here allows the desired gain asymmetry to be achieved by controlling the optical loss of graphene through electrical gating with ion gels\cite{kim2016direct,gao2015high}. 
%CH: ``active'' usually means non-linear , active tuning of the EPs can be achieved.

\subsection{Exceptional points in Fabry-P\'{e}rot cavity structures} 

The investigation of EPs in Fabry-P\'{e}rot (FP) cavity structures offers another route towards novel photonics-based technologies\cite {miri2019exceptional,tang2021room,wu2023chip,farhat2020pt,li2023highly,sakotic2023non,stone2020reflectionless,qin2020induced}. FP cavities, known for their ability to enhance optical resonances\cite {delbeke2002high,artar2009fabry}, expands the possibilities for controlling and manipulating light in reosnant cavities, particularly due to the divergent sensitivity characteristic of EPs\cite{ferise2022exceptional,khurgin2020exceptional}. 
Moreover, devices based on FP cavity structures can also simultaneously act as a laser and a coherent perfect absorber at EPs \cite{wu2023chip}, reminiscent of waveguide structures. This intriguing behavior arises from the mathematical bifurcation of the spectrum of an EP, which here corresponds to the bifurcation of a pair of poles and zeroes in the scattering matrix eigenvalues. Such behavior has been demonstrated in an integrated semiconductor resonator with active and passive regions\cite{wong2016lasing,zhu2018accessing}. 

The investigation of EPs in FP cavities has intensified lately. For instance, in a very recent combined experimental-computational study \cite{li2023highly}, an FP cavity has been used to facilitate the coupling between single-exciton and localized plasmon mode at an EP, thus demonstrating the potential advantageous of EPs in engineering light-matter interactions. 
%%CH: deleted your next sentence because you didn't cite more recent studies

% These recent studies underscore the growing significance of EPs in designing FP cavity structures. 
%They not only enhance our fundamental understanding of non-Hermitian physics but also hold promise for the development of innovative photonic devices and technologies that harness the unique characteristics of EPs in FP cavities. As such, the exploration of EPs within FP cavity structures remains a dynamic and vital area of research within the photonics field.

\section{Exceptional points in plasmonics}

\begin{figure}[h]
\includegraphics[width = 1\linewidth]{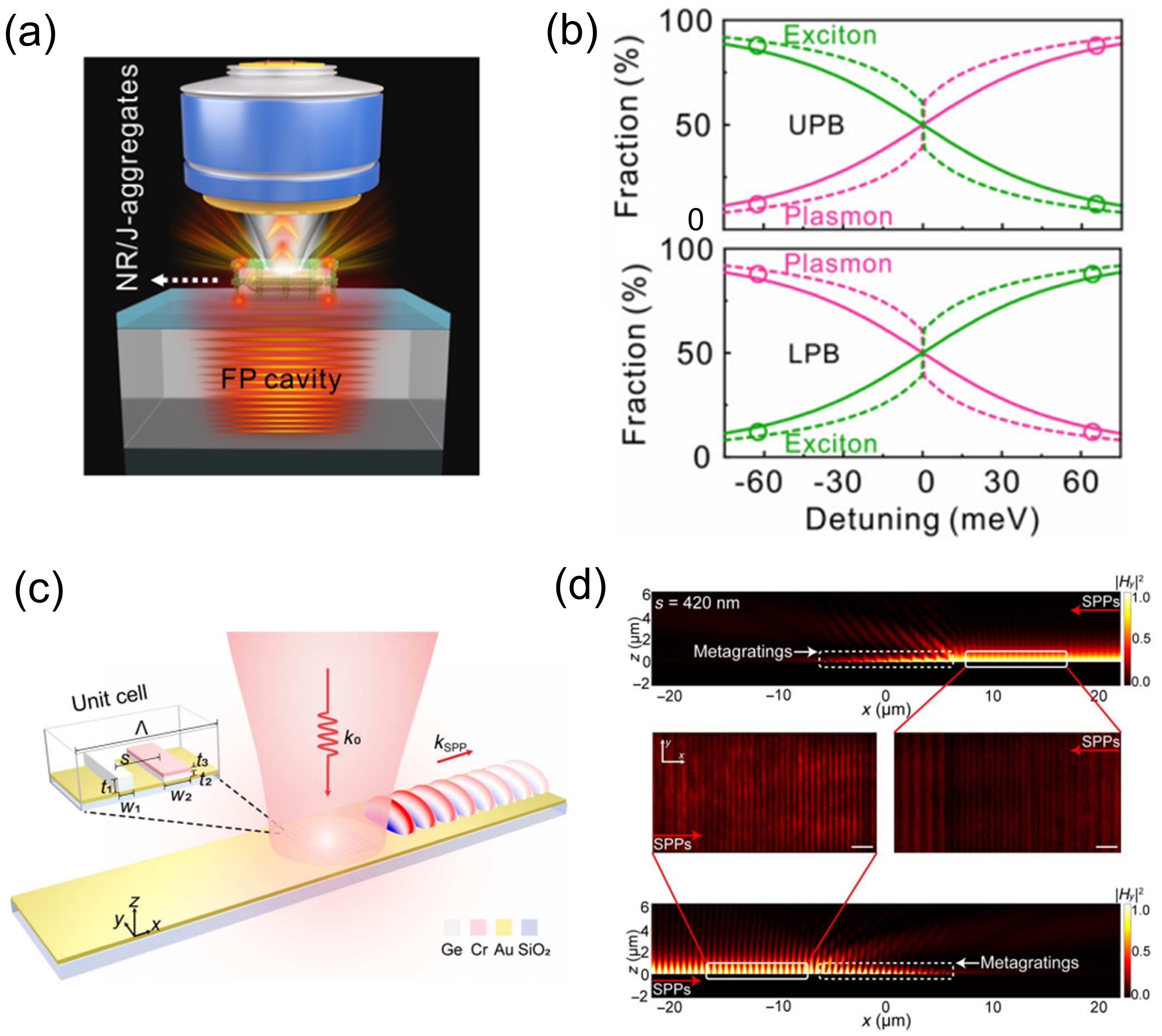}
\caption{ (a) Schematic of the experimental setup in Ref.~\onlinecite{li2023highly}, showing the Au@Ag nanorod/J-aggregate hybrid placed on a leaky Fabry-P\'{e}rot cavity. (b) Plasmonic and excitonic fractions in the upper/lower plexciton branch (UPB/LPB) for the single-exciton-coupled nanorods on the FP cavity (solid curves) and indium tin oxide substrate (dashed curves), respectively\cite{li2023highly}. (c) Schematic of the sub-wavelength metagrating setup in Ref.~\onlinecite{xu2023subwavelength}, which effectively corresponds to a spatially modulated permittivity with relative phase shift between its real and imaginary parts, a feature leading to unidirectional excitation of surface plasmon polaritons (SPPs) when tuned near an EP. (d) Simulated (top/bottom) and measured (middle) transverse magnetic field amplitudes $\left| H_y \right|^2$ for SPPs incident from the right (top) or left (bottom), showing asymmetrically much stronger interference amplitudes for left-incident SPPs\cite{xu2023subwavelength}. }
\label{fig:3}
\end{figure}

In the above, the waveguide and FP cavities that host EPs exist in the microscale regime, where classical electrodynamics dominate. Below, we describe recent developments in the study of EPs in plasmonic systems, which exist in the nanoscale regime where classical and quantum physics interplay\cite{butler2023exceptional,park2020symmetry,song2021plasmonic}, such that light and matter hybridize into new plasmonic and excitonic degrees of freedom. Plasmonic systems have gained much attention in the optics community due to their capacities in surpassing the diffraction limit, and enhancing light-matter interactions through surface plasmon polaritons \cite{pitarke2006theory, yu2019plasmon,zhou2022nanoscale,zhang2022coherent}. EPs in plasmonics highlight the nontrivial consequences of non-Hermitian $\mathcal{PT}$-symmetry breaking \cite{el2018non,ashida2020non} in the nanoscale regime, where the relevant physics are dictated by subwavelength-scale interactions between electromagnetic (EM) fields and the collective electronic oscillations in nanostructures. Furthermore, plasmonic resonances enable the concentrating electromagnetic fields at nanoscale length scales, thereby enhancing light-matter interactions. Therefore, devices such as sensors, lasers, and coherent perfect absorbers \cite{wiersig2020prospects,horstman2020exceptional,li2019exceptional,huang2015unidirectional,sakhdari2018low,jeong2023electrical,jiang2022exceptional,movsesyan2022engineering} can potentially be achieved by the synergies of plasmonics and non-Hermiticity.  EPs in plasmonic system can thus enable enhanced sensitivity to any permittivity perturbations, influencing resonance behavior and enabling fine-tuning of light-matter interactions. The integration of EPs withn plasmonic systems thus enhance the efficiency and sensitivity of plasmonic devices. 

A recent experiment\cite{li2023highly} demonstrated the novel application of EP physics in lowering the critical interaction strength needed for attaining strong single-exciton coupling with localized plasmons, with the setup schematically shown in Fig.~\ref{fig:3}a. This was achieved by suppressing the significant plasmon damping through a leaky Fabry-P\'{e}rot cavity, and matching the plasmonic and excitonic modes at an EP. As shown in Fig.~\ref{fig:3}b for the plasmonic (green) and excitonic (magenta) fractions in the upper/lower plexciton branches (UPB/PLB), shown on the top and bottom panels respectively, a sharp change in the fractions can be observed at resonance for the hybrid samples measured on the indium tin oxide substrate (dashed). This indicates that no level splitting occurs from the substrate, such that the observed spectral Rabi splitting is a pseudostrong coupling. In contrast, the corresponding fractions (solid) on the single-exciton-coupled nanorods do not exhibit such abrupt jumps due to the significant suppression of plasmonic damping. This breakthrough provides a highly efficient approach for achieving strong effective couplings between an exciton and plasmons by relaxing the effective threshold for strong coupling, thus mitigating the need for ultrasmall mode volume and precise dipole moment alignment \cite{liu2017strong,yang2022strong,tserkezis2020applicability}.

The synergy of plasmonics and EPs for enhancing and manipulating light-matter interactions has been demonstrated in another recent work \cite{xu2023subwavelength} in which subwavelength non-Hermitian metagratings are employed to enable unidirectional control of surface plasmon polaritons at the EPs [see Fig.\ref{fig:3}c]. The setup consists of a metal-dielectric interface where the dielectric layer has a spatially modulated permittivity profile with real and imaginary parts shifted by a relative phase, which is implemented in the physical setup through coupling with alternating nanostrips (white, pink).
%\begin{equation} 
%\begin{align}
%\epsilon(x)=\epsilon_d + A[cos\beta{x}-i V_0 sin(\beta{x}-\phi)] \\
%=\epsilon_d + A_L exp(i\beta{x})+A_R exp(-i\beta{x})
%\end{align}
%\end{equation} 
%$\epsilon_d$ is the background permittivity, V$_0$ is the relative modulation amplitude of the imaginary part of the permittivity, $\beta$ is the wave number of the surface plasmon polaritons, $\phi$ describes an additional phase shift of the imaginary modulation, and A represents the perturbation strength that is much smaller than $\epsilon_d$.The excitation of the surface plasmon polaritons can be modified as the separation distance s changes. Additionally,  through changing the parameters (V$_0$ and $\phi$), it can get insight into the behavior of non-Hermitian system in the parameter space. This can help conclude that A$_{R/L}$ contributes to the momentum compensation for the surface plasmon polaritons propagating to the right/left side (+x/−x direction).} 
In the basis of surface plasmon polariton (SPP) eigenmodes, the system can be approximately described by a dynamical 2-component operator\cite{longhi2018exceptional,wu2022direction} that mathematically predicts unidirectional propagation at an EP. This directionality is verified in numerical COMSOL simulations in the top/bottom panels of Fig~\ref{fig:3}d, which respectively show that SPPs incident from the right/left are suppressed/enhanced at a particular optimized set of system parameters. These results are also qualitatively corroborated experimental CCD snapshots (middle), where the interference patterns for left-incident SPPs are observed to be much stronger. Even slightly away from the EP in the presence of permittivity perturbations, which can frequently occur due to fabrication imperfection, the directional asymmetry of SPPs in the non-Hermitian metagrating remains robust. Such EP-enabled control mechanisms open the door for advanced optical devices with enhanced functionalities, promising breakthroughs in various fields where precise light manipulation is crucial. 

 In summary, the confluence of EPs and plasmonics holds immense potential in reshaping the landscape of subwavelength-scale optics and light-matter interactions. These developments are poised to impact a wide array of applications, encompassing telecommunications, sensing technologies, and advanced imaging applications, with the promise of advancing the boundaries of optical science and engineering\cite{park2020symmetry,jiang2022exceptional,kodigala2016exceptional}.

\section{Outlook}
The abovementioned photonic applications of EPs in coupled waveguides, Fabry-P\'{e}rot resonators, photonic crystals and plasmonics have mostly hinged on linear single-body EP physics, specifically in the multi-valuedness, algebraic divergence and defectiveness of its single-body band structure. However, many optical phenomena such as lasing are governed by dynamical equations that naturally exhibit EP bifurcations\cite{kominis2017spectral,kominis2018exceptional,zhiyenbayev2019enhanced}. With more precise control over optical nonlinearities in recent experiments\cite{ma2022nonlinear,jung2022thermal}, such as demonstrated in Ref.\onlinecite{dai2023non} with silicon microring arrays, optical nonlinearities intrinsic to many optical materials hold the promise of exhibiting new phenomena such as nonlinearity-induced topological phases\cite{gong2023topological,maczewsky2020nonlinearity,tuloup2020nonlinearity,kirsch2021nonlinear,hu2021nonlinear} and Floquet solitons\cite{mukherjee2020observation}, with the non-linearity itself controlling the extent of gain/loss in driving the phase transition\cite{lumer2013nonlinearly,miri2016nonlinearity}. These advances provides a new route towards new photonic applications involving fast optical switching and modulation and even quantum information communications\cite{minzioni2019roadmap}. 

With the rapid evolution of nanomaterials research, it would also be interesting to explore new EP phenomenology in photonic nanostructures based on emergent nanomaterials, such as 2D layered materials\cite{koppens2011graphene}, topological semimetals\cite{liu2020semimetals,meng2022terahertz} and phase change materials\cite{wuttig2017phase,meng2023port}, particularly to understand whether such materials can render unusual functionalities and tunability near the onset of an EP.

Another tantalizing new development beyond the EP band structure is the discovery of exceptional bound states\cite{lee2022exceptional}, which are special eigenstates that show up as robust isolated spectral resonances when a system containing an EP is confined both spatially and spectrally. %Specifically, within the subspace spanned by one (but not both) of the bands touching at an EP, exceptional bound states are 
Protected by the geometric defectiveness of the EP, they represent novel bound states that possess a completely different origin from the more well-known topological and non-Hermitian skin states. While they were touted as a source of negative entanglement entropy in the quantum context\cite{lee2022exceptional}, in classical systems, they can still exist as special resonances, as observed in a recent experiment with an electrical circuit platform derived from a parent EP model\cite{zou2023experimental}. What would be most interesting, particularly in photonics, is the generalization of exceptional bound states into 2D or beyond, where they would manifest as special resonant frequency bands that exist by virtue of the existence of a mathematical EP in a parent system. Suitably designed, such ``exceptional bound'' bands could not only lead to a new generation of bound states in the continuum (BICs), but could also lead to interesting new interplays with Stark, topological, nonlinearity-induced or non-Hermitian skin localization.

\maketitle

\begin{acknowledgements}
HYM acknowledges support by the Natural Science Foundation of Hunan Province (2023JJ40612). The authors acknowledge support from Singapore's Ministry of Education (MOE) Tier-II Grant (Proposal ID: T2EP50222-0008).

\end{acknowledgements}

\section*{Author Declarations}

\subsection*{Conflict of Interest}
\noindent The authors declare that there are no conflicts of interest.

%\subsection*{Author Contributions}
%\noindent H. Y. M. performed the calculation and analysis. H. Y. M. and Y. S. A. conceptualized and initiated the project. Y. S. A. and C. H. L. supervised the project. All authors contributed to the writing of this work.

\section*{Data Availability}
The data that supports the findings of this study are available from the corresponding author upon reasonable request.

\bibliography{Reference,references_rev}

\end{document}